\documentstyle[12pt]{article}
\begin{document}

\title{TOPOLOGICALLY\ NONTRIVIAL\ SECTORS\ OF\ THE\ MAXWELL\ FIELD\ THEORY\ ON\
ALGEBRAIC\ CURVES}
\author{Franco Ferrari \\
%EndAName
Dipartimento di Fisica, Universit\`a di Trento\\
38050 Povo (TN), Italy\\
and LPTHE \footnotemark \thanks{%
Laboratoire associ\'e No. 280 au CNRS}\\
{\it Universit\'e Pierre er Marie Curie--PARIS VI}\\
{\it Universit\'e Denis Diderot--Paris VII}\\
{\it Boite $126$, Tour 16, $1^{er}$ \'etage}\\
{\it 4 place Jussieu}\\
{\it F-75252 Paris CEDEX 05, FRANCE}}
\date{July 1996}
\maketitle
\vspace{-5.0in} \hfill{Preprint UTF 379, PAR-LPTHE 9630} \vspace{4.5in}
\begin{abstract}
In this paper the Maxwell field theory is considered on the $Z_n$ symmetric
algebraic curves. As a first result, a large family of nondegenerate metrics
is derived for general curves. This allows to treat many differential
equations arising in quantum mechanics and field theory on Riemann surfaces
as differential equations on the complex sphere. The examples of the scalar
fields and of an electron immersed in a constant magnetic field will be
briefly investigated. Finally, the case of the Maxwell equations on curves
with $Z_n$ group of automorphisms is studied in details. These curves are
particularly important because they cover the entire moduli space spanned by
the Riemann surfaces of genus $g\le 2$. The solutions of these equations
corresponding to nontrivial values of the first Chern class are explicitly
constructed.
\end{abstract}

\section{INTRODUCTION}

The two dimensional gauge field theories on a manifold and in particular on
Riemann surfaces have recently been considered by various authors \cite{tdym}%
, \cite{tdymi}. The abelian case is particularly interesting because of the
presence of topologically nontrivial sectors labelled by the nonzero values
of the first Chern class \cite{egh}. The problem of deriving topologically
nontrivial solutions of the Maxwell equations on a manifold is relevant in
string theories \cite{kl}, in Quantum Mechanics \cite{vo} and in the theory
of the Quantum Hall effect \cite{fqhe}. The classical solutions of the
Maxwell equations on Riemann surfaces have been derived in ref. \cite{ienli}
for particular metrics and in ref. \cite{ffunp} for conformally flat
metrics. More recently, the results of \cite{ffunp} have been generalized to
any metric in \cite{ffjpamg}. In the works mentioned above the gauge field
configurations corresponding to nonvanishing values of the first Chern class
have been expressed in terms of the prime form and theta functions \cite{fay}%
. In this paper, we use instead an alternative approach, which has already
provided physically relevant results in string theory for its explicitness 
\cite{acs}, \cite{acsi}. Namely, the Riemann surface will be represented
here as an $n-$sheeted covering of the complex sphere, i. e. as an affine
algebraic curve \cite{grihar}. One advantage of this representation is that
any differential equation defined on a Riemann surface, such as those
arising for instance in Quantum Field Theory or Quantum Mechanics, becomes a
differential equation on the sphere. A second advantage is that, at least in
the case of algebraic curves with $Z_n$ symmetry group of automorphisms, the
dependence on the moduli is explicit. The latter are in fact given by the
branch points and enter in the equation of the curve as simple complex
parameters. Despite of these advantages, the analytic solution of the
physically relevant field or wave equations is very difficult on an
algebraic curve because of the presence of multivalued differential
operators. Until now, only the theory of the socalled $b-c$ systems has been
fully solved on any algebraic curve using the operator formalism of refs. 
\cite{acs}--\cite{acsi}. The Maxwell field theory is more complicated than
the $b-c$ systems because it is not conformally invariant and the metric is
present in the equations of motion. Unfortunately, on an algebraic curve the
powerful tools offered by the theory of theta functions and exploited for
instance in refs. \cite{ienli} and \cite{ffjpamg}, are not very helpful for
computational purposes. For instance, the already existing formulas of the
prime form \cite{fay} are somewhat cumbersome and unexplicit for physical
applications.\\Despite of these problems, we will derive in this paper all
the nontrivial classical solutions of the Maxwell equations at least in the
case of the $Z_n$ symmetric curves. This is an important class of curves,
which covers the entire moduli space spanned by the Riemann surfaces of
genus $g\le 2$. We will see that the topologically nontrivial gauge fields
obtained here have a relatively simple expression, similar to their
analogues on the complex sphere and thus can provide new insights in two
dimensional Quantum Mechanics and FQHE on a manifold. Moreover, we derive a
big family of tensors describing nondegenerate metrics on the $Z_n$
symmetric curves. This allows to write explicitly in terms of multivalued
differential operators on the sphere not only the Maxwell equations, but
also any other equation of motion that involves particles of integer spin
living on a Riemann surface.\\Concerning the Maxwell equations, the greatest
difficulty in solving them is provided by the lack of a compact expression
for the prime form. In fact, apart from the case of the $C\theta M$ metric
of \cite{ienli}, the nontrivial gauge field configurations are expressed by
means of the prime form \cite{ffjpamg}. On the other side, on algebraic
curves it is difficult to construct the $C\theta M$ metric because the
canonically normalized differentials are not known (see for instance ref. 
\cite{ffv} for an attempt to construct these differentials). To circumvent
these difficulties, we give here particular kinds of nondegenerate metrics
which are singlevalued, i.e. they are invariant under arbitrary permutations
of the sheets composing the algebraic curve. Using these metrics the Maxwell
equations simplify considerably and the computation of their classical
solutions is made possible. \\The material presented in this paper is
divided as follows. In Section 2 the Maxwell field theory on Riemann
surfaces and algebraic curves is briefly introduced. The gauge fields are
decomposed in their exact, coexact and harmonic components using the Hodge
decomposition theorem. The harmonic components are explicitly derived. In
Section 3 we construct nondegenerate metrics on the $Z_n$ algebraic curves.
For some of these metrics the corresponding Ricci tensor is computed. We
check using the Poincar\'e--Lelong equation \cite{grihar}, that different
curvature tensors yield the same Euler characteristics as expected. In
Section 4 the topologically nontrivial solutions of the Maxwell equations
are derived. We verify that the magnetic fluxes generated by these gauge
fields satisfy the Dirac quantization condition. Finally, in the Conclusions
we discuss the possible applications and generalizations of our results. In
particular, it is shown how to generalize the metrics of Section 3 to any
affine algebraic curve. Moreover, the equations of motion of the scalar
fields on algebraic curves are treated with some details and the Hamiltonian
of a massive electron immersed in a constant magnetic field is explicitly
constructed.

\section{THE MAXWELL FIELD THEORY ON ALGEBRAIC CURVES}

In this paper we consider the Maxwell field theory on a Riemann surface $%
\Sigma $ of genus $h>1$ and with $Z_n$ group of symmetry. The action is
given by: 
\[
S_{Maxwell}=\int_\Sigma d\xi^1\wedge d\xi^2 \frac{\sqrt{g}}4F_{\mu \nu
}F^{\mu \nu } 
\]
where $F_{\mu \nu }=\partial _\mu A_\nu -\partial _\nu A_\mu $, $\mu ,\nu
=1,2$ and $\xi ^1,\xi ^2$ form a system of real coordinates on $\Sigma $.
Finally, $g_{\mu \nu }$ is the Euclidean metric on $\Sigma $ with
determinant $g$. The only nonvanishing components of the field strength are
given by $F_{21}=-F_{12}$. On a Riemann surface it is always possible to
assume that the metric $g_{\mu \nu }$ is conformally flat. In the latter
case, it is convenient to choose on $\Sigma$ complex coordinates $\xi =\xi
^1+i\xi ^2$ and $\bar \xi =\xi ^1-i\xi ^2$, so that the components of the
field strength ad of the metric become respectively: 
\[
F_{\xi \bar \xi }=-F_{\bar \xi \xi }=-{\frac 1{2i}}F_{12} 
\]
and 
\[
g_{\xi \bar \xi }=g_{\bar \xi \xi }={\frac{1}{2}} \sqrt{g}\qquad \qquad
\qquad g_{\xi \xi }=g_{\bar \xi \bar \xi }=0 
\]
Moreover, the volume form in complex coordinates is given by: $d^2\xi
g_{\xi\bar\xi}=d\xi^1\wedge d\xi^2\sqrt{g}$, where $d^2\xi\equiv id\bar
\xi\wedge d\xi$. Accordingly, the classical equations of motion of the
Maxwell field theory take the following form: 
\begin{eqnarray}
\partial _{\bar \xi }\left[ g^{\xi \bar \xi }\left( \partial _{\bar \xi
}A_\xi -\partial _\xi A_{\bar \xi }\right) \right] &=&0  \label{equmot} \\
\partial _\xi \left[ g^{\xi \bar \xi}\left( \partial _\xi A_{\bar \xi
}-\partial _{\bar \xi }A_\xi \right) \right] &=&0  \label{eqmot}
\end{eqnarray}
with $g^{\xi\bar \xi}$ being the inverse metric: $g^{\xi\bar \xi}g_{\xi\bar
\xi}=1$.

Let us decompose the gauge fields using the Hodge decomposition: 
\begin{equation}
A_\xi=\partial_\xi\varphi+\partial_\xi\rho+A^{{\rm har}}_\xi\qquad\qquad
A_{\bar \xi}=-\partial_{\bar \xi}\varphi+ \partial_{\bar \xi}\rho+A^{{\rm har%
}}_{\bar \xi}  \label{hodge}
\end{equation}
The coexact and exact components are expressed using the two scalar fields $%
\varphi$ and $\rho$ respectively. Here $\varphi$ is purely complex, while $%
\rho$ is real. $A^{{\rm har}}_\xi$ and $A^{{\rm har}}_{\bar \xi}$ take into
account the holomorphic differentials. The decomposition \ref{hodge} is not
invertible unless 
\begin{equation}
\int_\Sigma d^2\xi g_{\xi\bar \xi}\varphi(\xi,\bar \xi)= \int_\Sigma d^2\xi
g_{\xi\bar \xi}\rho(\xi,\bar \xi)=0  \label{consistency}
\end{equation}
The only nontrivial solutions of the equations \ref{equmot}--\ref{eqmot} are
provided by the $h$ harmonic differentials $A_{i,\xi }^{har}$, $A_{i,\bar
\xi }^{har}$, $i=1,\ldots ,h$ and by the vector fields $A_\xi ^I$, $A_{\bar
\xi }^I$ corresponding to nonvanishing values of the first Chern class. 
%\begin{equation}
The former satisfy the relations: 
\[
\partial _{\bar \xi }A_{i,\xi }^{har}=\partial _\xi A_{i,\bar \xi }^{har}=0 
\]
while the latter are builded in such a way that 
\[
F_{\xi \bar \xi }d\xi\wedge d\bar \xi= {\frac{i\Phi}A}g_{\xi \bar \xi
}d\bar\xi\wedge d\xi 
\]
where $\Phi$ is a constant representing the total magnetic flux associated
with the fields $A^I_\xi$, $A^I_{\bar \xi}$ and $A=i\int_\Sigma d^2\xi
g_{\xi \bar \xi }$ denotes the area of the Riemann surface. \\At this point
we represent $\Sigma $ in an explicit way as an algebraic curve associated
with Weierstrass polynomials of the kind: 
\begin{equation}
y^n=\prod_{i=1}^{nm}(z-a_i)  \label{zsymm}
\end{equation}
In \ref{zsymm}, $z$ and $\overline{z}$ denote a set of complex variables
describing the sphere {\bf CP}$_1$ and $n,m$ are integers. For our purposes,
it will be convenient to regard {\bf CP}$_1$ as the compactified complex
plane, i.e. {\bf CP}$_1\equiv ${\bf C}$\cup \left\{ \infty \right\} $. As
usual, we cover {\bf CP}$_1$ with two open sets $U_1$ and $U_2$ containing
the points $z=0$ and $z=\infty$ respectively. The local coordinate $%
z^{\prime}$ on $U_2$ is related to $z$ by the conformal transformation: $%
z^{\prime}={\frac{1}{z}}$. The $a_i$ are complex parameters denoting the
branch points of the curve. It is easy to check that the point at infinity $%
z=\infty $ is not a branch point. Solving eq. \ref{zsymm} with respect to $y$%
, we obtain a multivalued function $y(z)$, whose branches will be denoted
with the symbol $y^{(l)}(z)$, $l=0,\cdots ,n-1$. A generic tensor on the
algebraic curve can be multivalued on the complex sphere due to its
dependence on $y^{(l)}(z)$. To indicate the branches of any such tensor $T(z,%
\overline{z})$, we will use the following convenient notation:

\begin{equation}
T^{(l)}(z,\overline{z})\equiv T(z,\overline{z};y^{(l)}(z),\overline{y}^{(l)}(%
\overline{z}))  \label{tensnot}
\end{equation}

Any $Z_n$ symmetric algebraic curve $\Sigma $ is conformally equivalent to a
Riemann surface represented as a branched covering of {\bf CP}$_1$. The
genus of the surface is 
\begin{equation}
h=1-n+\frac{nm(n-1)}2  \label{genus}
\end{equation}
In practice, the Riemann surface $\Sigma $ is constructed gluing together
along a suitable set of branch lines $n$ copies of the complex sphere, the
so--called sheets. Accordingly, the $z$ variable can be viewed as a mapping $%
z:\xi\in\Sigma\rightarrow${\bf CP}$_1$. In this paper it will be always
understood that $z$ is a function of $\xi$, i.e. $z=z(\xi)$, unless
conversely stated. To construct a system of branch lines which is consistent
with the multivaluedness of the function $y(z)$, we group the branch points
in $m$ different sets $I_i=\{a_{(i-1)n+1},\cdots ,a_{in}\}$, with $%
i=1,\cdots ,m$. The branch points in a given set $I_i$ are connected
together by branch lines as shown in fig. 1. As a convention, going around a
branch point clockwise (counterclockwise) on the $j$--th sheet along a small
circle surrounding the point, one encounters the $(j+1)$--th ($(j-1)$--th)
sheet when crossing a branch line, where $j=0,\cdots ,n-1\;{\rm {mod}\,n}$.

In order to describe the topologically nontrivial solutions of the Maxwell
field theory on the algebraic curves \ref{zsymm} explicitly, the following
relevant divisors are necessary \footnotemark {}
\footnotetext{
These divisors can be computed using the methods of refs. \cite{acsi} and 
\cite{ffi}.}: 
\begin{eqnarray}
\left[ dz\right] &=&(n-1)\sum_{p=1}^{nm}a_p-2\sum_{j=0}^{n-1}\infty _j 
\nonumber \\
\left[ y\right] &=&\sum_{p=1}^{nm}a_p-m\sum_{j=0}^{n-1}\infty _j
\label{divisors}
\end{eqnarray}
In \ref{divisors} the symbol $\infty _j$ denotes the projection of the point 
$z=\infty $ on the $j$--th sheet. Exploiting the above divisors, it is easy
to see that the holomorphic differentials $A_{i,\xi }^{har}$ and $A_{i,\bar
\xi }^{har}$ correspond to linear combinations of the following holomorphic
differentials: 
\begin{equation}
\Omega _{k,j}dz={\frac{z^{j-1}}{y^{-k+n-1}}}dz  \label{holodiffs}
\end{equation}
where 
\[
\left\{ 
\begin{tabular}{c}
$j=1,\cdots ,(n-1)m-km-1$ \\ 
$k=0$,$\cdots ,n-2$%
\end{tabular}
\right. 
\]
for $m>1$ and 
\[
\left\{ 
\begin{tabular}{c}
$j=1,\cdots ,n-k-2$ \\ 
$k=0,\cdots ,n-3$%
\end{tabular}
\right. 
\]
for $m=1$.

The calculation of the topologically nontrivial solutions $A_\xi ^I$ is more
complicated, for we need an explicit expression of the metric on $\Sigma $.
This will be the subject of the next Section.

\section{METRIC TENSORS ON ALGEBRAIC CURVES}

A simple class of conformally flat metrics is provided by tensors of the
kind: 
\begin{equation}
g_{z\overline{z}}dzd\overline{z}=\frac{dzd\overline{z}}{\left( y\overline{y}%
\right) ^{n-1}}\left[ 1+f(z,y)\overline{f(z,y)}\right] ^\alpha
\label{mgenmetr}
\end{equation}
where $f(z,y)$ is a rational function of $z$ and $y$ and $\overline{f(z,y)}$
its complex conjugate. Let us notice that the parameter $\alpha $ can also
be a rational number. However, it is clear that the term $\left[ 1+f(z,y)%
\overline{f(z,y)}\right] ^\alpha $ does not introduce further branches on 
{\bf CP}$_1$. Putting $\alpha =0$ in equation \ref{mgenmetr}, we obtain the
degenerate metric

\[
g_{1,z\overline{z}}dzd\overline{z}=\frac{dzd\overline{z}}{\left( y\overline{y%
}\right) ^{n-1}} 
\]

which has already been used in perturbative string theory. For $\alpha =%
\frac{(n-1)m-2}m$ and $f(z,y)=y(z)$, we have instead the nondegenerate
metric 
\begin{equation}
g_{2,z\overline{z}}dzd\overline{z}=\frac{dzd\overline{z}}{\left( y\overline{y%
}\right) ^{n-1}}\left[ 1+y\overline{y}\right] ^\alpha  \label{nondegenerate}
\end{equation}

To compute the Ricci tensor $R_{2,z\overline{z}}$ corresponding to the above
conformally flat metric, it is convenient to put 
\[
\frac 12{\rm e}^{2\sigma }=\frac{(1+y\overline{y})^\alpha }{\left( y%
\overline{y}\right) ^{n-1}} 
\]
In terms of $\sigma $, $R_{2,z\overline{z}}$ is defined as follows: 
\[
R_{2,z\overline{z}}=-8\partial _z\partial _{\overline{z}}{\rm \sigma } 
\]

In the explicit calculation of $R_{2,z\overline{z}}$, the contributions to
the total curvature coming from the term $\partial _z\partial _{\overline{z}}%
{\rm \log }\left[ y\right] $ and its complex conjugate are proportional to
Dirac $\delta $--functions concentrated at the branch points and at
infinity. However, these $\delta $--functions cancel in the final expression
of the Ricci tensor because the metric $g_{2,z\overline{z}}$ is
nondegenerate and so it does not have poles or zeros at those points. Thus,
after a straightforward calculation, we obtain:

\begin{equation}
R_{2,z\overline{z}}=-4\alpha \frac{\partial _zy\partial _{\overline{z}}%
\overline{y}}{\left[ 1+y\overline{y}\right] ^2}  \label{curvature}
\end{equation}
Another example of nondegenerate metric is provided by: 
\begin{equation}
g_{3,z\overline{z}}dzd\overline{z}=\frac{dzd\overline{z}}{\left( y\overline{y%
}\right) ^{n-1}}\left[ 1+z\overline{z}\right] ^\beta  \label{nondegtwo}
\end{equation}
for $\beta =(n-1)m-2$. The Ricci tensor corresponding to this metric has a
very simple form: 
\[
R_{3,z\overline{z}}=-\frac{4\beta }{\left( 1+z\overline{z}\right) ^2} 
\]
To verify that the above tensors represent real metrics on $\Sigma$, we will
compute the Euler characteristic $\chi $ for the $Z_n$ symmetric algebraic
curves. On a Riemann surface $M_g$ of genus $g$ equipped with a metric $%
G_{\mu \nu }$ and global complex coordinates $\xi $ and $\overline{\xi }$, $%
\chi $ is defined as follows: 
\[
\chi =\frac 1{4\pi }\int_{M_g}d^2\xi \sqrt{G}R 
\]
where $R$ is the curvature scalar and $d^2\xi\sqrt{G}=\sqrt{G}d\xi\wedge
\bar\xi$ is the volume form. As already mentioned, in the representation of
Riemann surfaces in terms of algebraic curves, $z$ can be viewed as the
mapping $z:\xi \in \Sigma \rightarrow ${\bf CP}$_1$. Therefore, the integral
of a density $L_{z\overline{z}}^{(l)}(z,\overline{z})$ over $\Sigma$
becomes: 
\[
I=\int_\Sigma d^2z(\xi )L_{z\overline{z}}^{(l)}(z(\xi ),\overline{z}(%
\overline{\xi })) 
\]
To evaluate integrals of this kind, it is possible to use the
Poicar\'e--Lelong equation \cite{grihar}, which expresses $I$ as an integral
over {\bf CP}$_1\otimes ${\bf CP}$_1$: 
\begin{equation}
I=\frac i\pi \int_{{\bf CP}_1}d^2z\int_{{\bf CP}_1}d^2y\partial _y\partial _{%
\overline{y}}L_{z\overline{z}}(z,\overline{z};y,\overline{y}){\rm \log }%
\left| F(z,y)\right|  \label{plelong}
\end{equation}
where $F(z,y)=y^n-\prod_{i=1}^{nm}(z-a_i)$. After an integration by parts in 
$\partial_y$ and $\partial_{\bar y}$, the latter formula simplifies to 
\begin{equation}
I=\sum_{l=0}^{n-1}\int_{{\bf CP}_1}d^2zL_{z\overline{z}}^{(l)}(z,\overline{z}%
)  \label{tplelong}
\end{equation}
This form of the Poincar\'e--Lelong equation is particularly convenient for
our aims. It can be derived as in \cite{grihar} exploiting the following
Cauchy formula: 
\[
\frac 1{2\pi i}\partial _y\partial _{\overline{y}}\log \left|
y-y^{(l)}(z)\right| =\delta _{y\overline{y}}^{(2)}(y,y^{(l)}(z)) 
\]
which is a straightforward consequence of the fact that $F(z,y)$ can be
rewritten as follows: $F(z,y)=\prod_{l=0}^n(y-y^{(l)}(z))$. Equivalently, if
we consider $y$ as the independent variable instead of $z$, so that $z(y)$
becomes a multivalued function with $s=1,\cdots ,nm$ branches, we have: 
\begin{equation}
I=\sum_{s=0}^{nm-1}\int_{{\bf CP}_1}d^2yL_{z\overline{z}}^{(s)}(y,\overline{y%
})  \label{fplelong}
\end{equation}
In the case of the Euler characteristic, taking into account for instance
the metric $g_{3,z\bar z}$ with Ricci tensor $R_{3,z\bar z}$, the relevant
integral to be computed is: 
\begin{equation}
\chi =-\frac 1{4\pi }\int_{{\bf \Sigma }}d^2z(\xi )\frac{4\beta }{\left(
1+z(\xi )\overline{z}(\overline{\xi })\right) ^2}  \label{glad}
\end{equation}
Since the integrand in the above equation is not multivalued and $z$ maps
the curve into $n$ copies of {\bf CP}$_1$, it is easy to realize that the
right hand side of \ref{glad} is equivalent to the following integral over 
{\bf CP}$_1$: 
\begin{equation}
\chi =-\frac n{4\pi }\int_{{\bf CP}_1}d^2z\frac{4\beta }{\left( 1+z\overline{%
z}\right) ^2}  \label{glada}
\end{equation}
This formula can also be directly verified using eq. \ref{tplelong}. At this
point it is sufficient to notice that the right hand side of \ref{glada} is
proportional to the Euler characteristic of the complex sphere: 
\begin{equation}
\chi _{{\bf CP}_1}=\frac 1\pi \int_{{\bf CP}_1}\frac{d^2z}{\left( 1+z%
\overline{z}\right) ^2}=2  \label{chisphere}
\end{equation}
Therefore, eq. \ref{glada} yields $\chi =-4n\beta =(2n-nm(n-1))$. Comparing
this value of $\chi $ with eq. \ref{genus}, which gives the genus of $\Sigma 
$ in terms of $n$ and $m$, it is easy to see that $\chi =2-2g$ as expected.

The computation of the Euler characteristic starting from the Ricci tensor $%
R_{2,z\bar z}$ of eq. \ref{curvature} can be performed in an analogous way.
In this case $\chi $ is given by 
\begin{equation}
\chi =-4\alpha \int_\Sigma d^2z(\xi )\frac{\partial _zy\partial _{\overline{z%
}}\overline{y}}{\left[ 1+y\overline{y}\right] ^2}  \label{czz}
\end{equation}
To deal with the above integral it is convenient to consider $y$ as the
independent variable and to solve eq. \ref{zsymm} with respect to $z$. Thus $%
y$ maps $\Sigma $ into $nm$ copies of {\bf CP}$_1$. With the help of eq. \ref
{fplelong}, eq. \ref{czz} becomes after a straightforward calculation: 
\[
\chi =-4\alpha \int_\Sigma \frac{d^2y(\xi )}{\left[ 1+y(\xi )\overline{y}(%
\overline{\xi })\right] ^2}=-4\alpha nm\int_{{\bf CP}_1}\frac{d^2y(\xi )}{%
\left[ 1+y(\xi )\overline{y}(\overline{\xi })\right] ^2} 
\]
Exploiting again eq. \ref{chisphere}, we obtain the correct value of the
Euler characteristic on a Riemann surface: $\chi =2-2g$.

Let us finally notice that we can construct other metric tensors on an
algebraic curve which are not of the form \ref{mgenmetr}. For instance the
tensor 
\begin{equation}
\widetilde{g}_{z\overline{z}}dzd\overline{z}=\frac{e^{\left( y\overline{y}%
\right) ^{n-1}}}{e^{\left( y\overline{y}\right) ^{n-1}}-1}\left[ 1+z%
\overline{z}\right] ^{\frac 2m}dzd\overline{z}  \label{nondegthree}
\end{equation}
yields a nondegenerate metric, as it is easy to check exploiting the
divisors \ref{divisors}. Moreover, all the metric tensors given above are
characterized by the fact that they are not multivalued on {\bf CP}$_1$. In
fact, they depend on $y$ only through the modulus of this function, which is
branch independent. Nondegenerate metrics which are also multivalued can be
for instance obtained multiplying eqs. \ref{nondegenerate}, \ref{nondegtwo}
and \ref{nondegthree} by integer powers of the factor $\frac y{\overline{y}}+%
\frac{\overline{y}}y$ or by considering more complicated forms of the
functions $f(z,y)$ in eq. \ref{mgenmetr}

%Finally, on a general algebraic curve with Weierstrass
%polynomial $F(z,y)$ the metrics given here can be easily generalized by
%substituting the 

\section{SOLITONIC SECTORS OF THE MAXWELL FIELD THEORY}

At this point, we are ready to derive the fields $A_\xi^I$ and $A_{\bar\xi}^I
$ explicitly. On the algebraic curve, this is equivalent to solve the
equation: 
\begin{equation}
F_{z\bar z}dz\wedge d\bar z={\frac{i\Phi}{A}}g_{z\bar z}dz\wedge d\bar z
\label{fund}
\end{equation}
The difficulty of computing $A_\xi^I$ and $A_{\bar \xi}^I$ explicitly
strongly depends on the choice of the metric $g_{z\bar z}$. For instance, in
the formalism of theta functions, eq. \ref{fund} can be easily solved in the
canonical $\theta$--metric (C$\theta$M) of \cite{ienli}. However, in order
to construct the C$\theta$M metric, we would need at least the explicit
expression of the period matrix, which is not known on an algebraic curve.
For these reasons, we will choose here metrics $g_{z\bar z}$ which are
singlevalued on {\bf CP}$_1$. Examples of nondegenerate metric tensors of
this kind are provided by eqs. \ref{nondegenerate}, \ref{nondegtwo} and \ref
{nondegthree}. To solve eq. \ref{fund}, we first define the following Green
function: 
\[
G(z,w)=-{\frac{1}{4\pi n}}{\rm log}\left[{\frac{|z-w|^2}{(1+z\bar z)
(1+w\bar w)}}\right]
\]
$G(z,w)$ is proportional to the inverse of the Laplace operator on {\bf CP}$%
_1$ and it is a well defined Green function also on $\Sigma$, where it
satisfies the equation 
\[
\partial_z\partial_{\bar z}G(z,w)=-{\frac{1}{2n}}\delta^{(2)}_{z\bar z}(z,w)
+{\frac{\gamma_{z\bar z}}{4\pi nA}}
\]
$\delta^{(2)}_{z\bar z}(z,w)$ is formally the usual Dirac $\delta$--function
on the complex sphere, but we have to remember that on the algebraic curve $%
\delta^{(2)}_{z\bar z}(z,w)$ is singular at all the projections of the point 
$z=w$ on $\Sigma$.

Let us denote with $J(z,\bar z)$ an external singlevalued scalar current on 
{\bf CP}$_1$. Since $g_{z\bar z}$ is singlevalued, it is easy to show with
the help of the Poincar\'e--Lelong formula that the following result is
valid: 
\[
\partial_z\partial_{\bar z}\int_\Sigma d^2
w(\xi^{\prime})G(w(\xi^{\prime}),z(\xi)) J(w(\xi^{\prime}),\bar w(\bar
\xi^{\prime}))g_{w\bar w}=-{\frac{J(z,\bar z)}{2}}g_{z\bar z}+ 
\]
\begin{equation}
{\frac{ \gamma_{z\bar z}}{4\pi n}}\int_\Sigma d^2w J(w,\bar w)g_{w\bar w}
\label{utile}
\end{equation}
where $\gamma_{z\bar z}={\frac{1}{(1+z\bar z)^2}}$ is the standard metric on
the complex sphere. We notice that $\gamma_{z\bar z}$ could be taken as a
metric on $\Sigma$, but unfortunately it is degenerate because of its zeros
at the branch points (see eq. \ref{divisors}). However, this fact will not
be disturbing in the following calculations, since $\gamma_{z\bar z}$ will
only appear as an auxiliary tensor.

Besides the Green function $G(z,w)$, we also introduce a gauge field $%
A_\alpha^{sph}$, $\alpha=z,\bar z$, defined in this way: 
\begin{eqnarray}
A_z^{sph}&=-{\frac{A}{4\pi n}}\partial_z{\rm log}(1+z\bar z)  \label{auxone}
\\
A_{\bar z}^{sph}&={\frac{A}{4\pi n}}\partial_{\bar z}{\rm log}(1+z\bar z)
\label{auxtwo} \\
\end{eqnarray}
It is easy to check that the following relation 
\begin{equation}
\partial_z A_{\bar z}^{sph}-\partial_{\bar z}A_z^{sph}={\frac{A\gamma_{z\bar
z} }{4\pi n}}  \label{local}
\end{equation}
is satisfied over all the algebraic curve $\Sigma$ apart from the points $%
\infty_0,\ldots,\infty_{n-1}$. At those points, in fact, a $\delta$%
--function concentrated in $z=\infty$ appears in the right hand side of \ref
{local}. To show this, it is just sufficient to perform the change of
variables $z^{\prime}=1/z$ in eqs. \ref{auxone}--\ref{auxtwo} and to study
the behavior of the right hand side in $z^{\prime}=0$. The problem of the
appearance of $\delta$ functions can be solved as in the case of the
Wu--Yang monopoles on the sphere by splitting the algebraic curve into two
sets $\Sigma_S$ and $\Sigma_N$. The former contains all the projections of
the point $z=0$ but not those of the point $z=\infty$, while for $\Sigma_N$
the converse is true. Of course, there is a great arbitrariness in choosing
the sets $\Sigma_S$ and $\Sigma_N$. To fix the ideas, we will define them as
the two sets obtained by cutting the algebraic curve along the contour $%
\gamma$ shown in fig. 2. Thus $\Sigma_S$ encloses the projections $%
0_0,\ldots,0_{n-1}$ of the point $z=0$ and all the branch points apart from $%
a_{nm}$. Consequently, $\Sigma_N$ includes the points $\infty_0,\ldots,
\infty_{n-1}$ and the branch point $a_{nm}$. The contour of fig. 2 is valid
also in the case in which $z=0$ is a branch point. In fact we can always put 
$a_1=0$, without any loss of generality. On a two sphere, the above
decomposition corresponds to the usual decomposition in southern and
northern hemisphere. The difference in the present case is that $\Sigma_S$
and $\Sigma_N$ are not isomorphic to {\bf C}. Despite of that, this way of
covering the algebraic curve will be sufficient for our purposes as we will
show below. Indeed, let us write the solution of eq. \ref{fund} on $\Sigma_S$%
: 
\begin{equation}
A_z^S={\frac{i\Phi}{A}}\left[\int_\Sigma d^2w\partial_zG(z,w)g_{w\bar w}+
A_z^{sph}\right]  \label{sols}
\end{equation}
\begin{equation}
A_{\bar z}^S={\frac{i\Phi}{A}}\left[-\int_\Sigma d^2w\partial_{\bar z}
G(z,w)g_{w\bar w}+ A_{\bar z}^{sph}\right]  \label{solls}
\end{equation}
Exploiting eqs. \ref{utile} and \ref{local}, it turns out that: 
\begin{equation}
\left( \partial_z A_{\bar z}^S-\partial_{\bar z}A_z^S\right)dz\wedge d\bar z
={\frac{i\Phi}{A}} g_{z\bar z }d\bar z \wedge dz  \label{sres}
\end{equation}
as desired. Analogous expressions of the solutions of equation \ref{fund}
can be written on $\Sigma_N$. Using the coordinates $z^{\prime}=1/z$ and $%
\bar z^{\prime}=1/\bar z$ we have: 
\begin{equation}
A_{z^{\prime}}^N={\frac{i\Phi}{A}} \left[\int_\Sigma
d^2w^{\prime}\partial_{z^{\prime}}G(z^{\prime},w^{\prime})g_{w^{\prime}\bar
w^{\prime}}+ \tilde A_{z^{\prime}}^{sph}\right]  \label{soln}
\end{equation}
\begin{equation}
A_{\bar z^{\prime}}^N={\frac{i\Phi}{A}}\left[-\int_\Sigma
d^2w^{\prime}\partial_{\bar z^{\prime}}
G(z^{\prime},w^{\prime})g_{w^{\prime}\bar w^{\prime}}+ \tilde A_{\bar
z^{\prime}}^{sph}\right]  \label{solln}
\end{equation}
where 
\[
G(z^{\prime},w^{\prime})=-{\frac{1}{4\pi n}}{\rm log}\left[{\frac{%
|z^{\prime}-w^{\prime}|^2}{(1+z^{\prime}\bar z^{\prime}) (1+w^{\prime}\bar
w^{\prime})}}\right]
\]
and 
\begin{eqnarray}
\tilde A_{z^{\prime}}^{sph}&=- {\frac{A}{4\pi n}}\partial_{z^{\prime}}{\rm %
log}(1+z^{\prime}\bar z^{\prime})  \label{nauxone} \\
\tilde A_{\bar z^{\prime}}^{sph}&= {\frac{A}{4\pi n}}\partial_{\bar
z^{\prime}}{\rm log}(1+z^{\prime}\bar z^{\prime})  \label{nauxtwo} \\
\end{eqnarray}
As for the fields $A^{S}_{z}$ and $A^{S}_{\bar z}$, it is easy to prove that
the following relations are satisfied: 
\begin{equation}
\partial_{z^{\prime}} \tilde A_{\bar z^{\prime}}^{sph}-\partial_{\bar
z^{\prime}}\tilde A_{z^{\prime}}^{sph}= {\frac{A\gamma_{z^{\prime}\bar
z^{\prime}} }{4\pi n}}  \label{nlocal}
\end{equation}
and 
\begin{equation}
\left( \partial_{z^{\prime}} A_{\bar z^{\prime}}^N-\partial_{\bar
z^{\prime}}A_{z^{\prime}}^N\right)= {\frac{i\Phi}{A}} g_{z^{\prime}\bar
z^{\prime}}d\bar z^{\prime}\wedge dz^{\prime}  \label{nres}
\end{equation}
Equations \ref{sols}--\ref{solls} and \ref{soln}--\ref{solln} show the
reasons for which we only need two sets to cover the Riemann surface $\Sigma$%
. First of all, the gauge fields $A^N_\alpha$ and $A^S_\alpha$ with $%
\alpha=z,\bar z$, are singlevalued on $\Sigma$. Secondly, they are
everywhere well defined, i.e. free of singularities, apart from the
projections on the algebraic curve of the points $z=0,\infty$. As a
consequence, the behavior of the gauge fields $A^{N,S}_\alpha$ is not
affected by the presence of the branch points and the splitting of $\Sigma$
into two sets $\Sigma_N$ and $\Sigma_S$ is justified.\\To be consistent,
both fields $A^N_\alpha$ and $A^S_\alpha$ should describe the same magnetic
field. Indeed, far from the points $z=0$ and $z=\infty$, where it is
possible to use on any sheet of $\Sigma$ both coordinates $z(\xi)$ and $%
z^{\prime}(\xi)$, it is easy to see that the fields $A^{N,S}_\alpha$ are
related by a gauge transformation. Introducing the variable $\Lambda={\frac{%
i\Phi}{4\pi n}}{\rm log}{\frac{z^{\prime}}{\bar z^{\prime}}} $ we obtain in
fact the following relations: 
\begin{eqnarray}
A_{z^{\prime}}^Sdz^{\prime}=&A_{z^{\prime}}^Ndz^{\prime}+\partial_{z^{%
\prime}}\Lambda dz^{\prime}  \label{gone} \\
A_{\bar z^{\prime}}^Sd\bar z^{\prime}=&A_{\bar z^{\prime}}^Nd\bar
z^{\prime}+\partial_{\bar z^{\prime}}\Lambda d\bar z^{\prime}  \label{gtwo}
\\
\end{eqnarray}
Using local polar coordinates $z^{\prime}={\frac{1}{\rho^{\prime}}}%
e^{-i\theta^{\prime}}$, we have that 
\[
\Lambda={\frac{\Phi\theta^{\prime}}{2\pi n}}
\]
It is now easy to see that the transformations \ref{gone}--\ref{gtwo} can be
reabsorbed by an $U(1)$ gauge transformation corresponding to a group
element $U(z^{\prime},\bar z^{\prime})$ given by: 
\begin{equation}
U=e^{i\Lambda}=e^{i{\frac{\Phi\theta^{\prime}}{2\pi n}}}  \label{gelem}
\end{equation}
An analogous result can be found working in $z$ coordinates on $\Sigma_S$.
Since the fields $A^N_\alpha$ and $A^S_\alpha$ differ by an exact
differential, the corresponding field strength $F_{z\bar z}$ is globally
defined on $\Sigma$ and we can compute the total magnetic flux associated to
the gauge field configurations of eqs. \ref{sols}--\ref{solls} and \ref{soln}%
--\ref{solln}. Using eqs. \ref{sres} and \ref{nres} we have: 
\[
\int_\Sigma d^2zF_{z\bar z}^I=\int_{\Sigma_S}d^2zF_{z\bar z}^S+
\int_{\Sigma_N}d^2z^{\prime}F_{z^{\prime}\bar z^{\prime}}^N=\Phi 
\]
as desired, where $F^I_{z\bar z}=\partial_z A^I_{\bar z}- \partial_{\bar
z}A^I_z$ and $I=N,S$ on $\Sigma_{N,S}$.\\Finally, let us check the
conditions for which the group element $U=e^{i{\frac{\Phi\theta^{\prime}}{%
2\pi n}}}$ of eq. \ref{gelem} is well defined, i.e. is singlevalued in the
intersection between $\Sigma_S$ and $\Sigma_N$ along the path $\gamma$. When 
$U$ is transported $N$ times along $\gamma$, the angle $\theta^{\prime}$
undergoes the following shift: 
\[
\theta^{\prime}\rightarrow \theta^{\prime}+2\pi n N
\]
as the contour $\gamma$ encircles all the projections of the point $z=\infty$
on the $n$ different sheets composing the algebraic curve. In order to
ensure singlevaluedness, the following condition on the total flux $\Phi$
should be imposed: 
\[
\Phi=2\pi k\qquad\qquad\qquad k=0,\pm 1,\pm 2,\ldots
\]
As a consequence, the solutions of eq. \ref{fund} provided by the gauge
field configurations in \ref{sols}--\ref{solls} and \ref{soln}--\ref{solln}
satisfy the relation: 
\begin{equation}
2\pi k=\int_\Sigma d^2z F_{z\bar z}  \label{fcc}
\end{equation}
with integer values of $k$. Thus, as expected, we recover exactly the Dirac
quantization of the flux. This result does not depend on the form of the
contour $\gamma$. A curve which encircles all the projections of the points $%
z=0$ and $z=\infty$ has either to cross at least $n$ branch lines as in fig.
2 or to be of the form given in fig. 3. In both cases, the total shift in
the angle $\theta^{\prime}$ has an $n$ factor in front which does not allow
for fractional values of $k$ in eq. \ref{fcc}. Also the situation in which
the point $z=0$ is a branch point does not represent a great complication 
\footnotemark {}
\footnotetext{
The case of a branch point at infinity can be treated in the same way of the
branch point at $z=0$. We will not discuss it here because branch points at
infinity are impossible due to the form of the Weierstrass polynomial \ref
{zsymm}.}. Indeed, a contour encircling a branch point must be also of the
kind given in figs. 2 and 3. In the latter case, one has to suppose that the
paths which surround the branch point $z=0$ on each sheet are builded in
such a way to avoid any branch line, including those outgoing from the point 
$z=0$. Moreover, if the contour $\gamma$ is very close to the branch point $%
z=0$, we can use local uniformization coordinates $z=t^n$ and $\bar z=\bar
t^n$. In these coordinates, assuming polar coordinates $t=\rho e^\theta$,
the group element $U$ of eq. \ref{gelem} becomes: 
\[
U=e^{i{\frac{\Phi\theta}{2\pi n}}}
\]
When $U$ is transported around the branch point $N$ times, the variable $%
\theta$ shifts as follows: $\theta\rightarrow \theta+2\pi N$. This gives
again the Dirac quantization condition \ref{fcc} of the total flux $\Phi$.
Since the gauge fields defined in eqs. \ref{sols}--\ref{solls} and \ref{soln}%
--\ref{solln} are singlevalued on {\bf CP}$_1$, no problem arises when they
are transported along the homology cycles.

\section{CONCLUSIONS}

In eqs. \ref{holodiffs} and \ref{sols}--\ref{solln} we have computed all the
nontrivial solutions of the Maxwell equations on a $Z_n$ symmetric algebraic
curve. The gauge fields configurations \ref{sols}--\ref{solls} and \ref{soln}%
--\ref{solln} have a simple form, as they are expressed through the scalar
Green function on the complex sphere and are singlevalued. Using these gauge
potentials it is possible to write explicitly the Hamiltonian $H$ of an
electron of mass $m$ in the presence of a constant magnetic field
perpendicular to $\Sigma$. Remembering that $B={\frac{\Phi}{A}}$, we have on 
$\Sigma_{N,S}$: 
\begin{equation}
H=[2g^{z\bar z}/m](P_z-A^{N,S}_z)(P_{\bar z}-A^{N,S}_{\bar z})+[\Phi/2mA]
\end{equation}
with $P_\alpha=-i\partial_\alpha$, $\alpha=z,\bar z$. $g^{z\bar z}$ is the
inverse of one of the singlevalued metrics given in Section 3. The
(degenerate) ground state of $H$ is given by 
\[
\Psi_{N,S}=e^{-\int_{\bar z_0}^{\bar z}A_{\bar z}^{N,S}d\bar z}\Psi_0
\]
where $\Psi_0$ satisfies the relation: $\partial_z\Psi_0=0$. Due to the
Dirac quantization condition of the magnetic flux and the singlevaluedness
of $A^{N,S}_\alpha$, $\Psi_{N,S}$ is a well defined quantum mechanical
state. In particular, it is not periodic along the homology cycles. In fact,
because of the singlevaluedness, the periods of the gauge potentials along
any closed curve $C$ defined on $\Sigma$ are zero: 
\[
\int_{C_N}A_{\bar z}^Nd\bar z+\int_{C_S}A^S_{\bar z}d\bar z=0
\]
Here $C_N$ and $C_S$ are the components of $C$ lying on $\Sigma_N$ and $%
\Sigma_S$ respectively. The way in which the topologically notrivial
solutions have been obtained here makes use of the $Z_n$ symmetry of the
curves. As a matter of fact, the singularities in $z=0$ and $z=\infty$ of
the gauge potentials \ref{sols}--\ref{solls} and \ref{soln}--\ref{solln}
appear symmetrically over all sheets of $\Sigma$. Therefore, our procedure
is not suitable for general algebraic curves. However, the metrics given in
Section 3 can be easily extended to any algebraic curve with Weierstrass
polynomial: 
\begin{equation}
F(z,y)=\sum_{i=0}^n y^iP_i(z)  \label{wpol}
\end{equation}
where the $P_i(z)$ are polynomials in $z$. In this case, the metrics \ref
{mgenmetr} takes the form: 
\[
g_{z\bar z}dzd\bar z {\frac{dzd\bar z}{|F_y|^2}}\left(1+f(z,y)\overline{%
f(z,y)} \right)^\alpha
\]
with $F_y(z,y)=\partial_yF(z,y)$. The values of the parameter $\alpha$
depend on the form of the polynomial \ref{wpol} and of the function $f(z,y)$%
. To determine $\alpha$, one has to derive the divisors of $dz$, $y$ and $F_y
$ as in eq. \ref{divisors}. For a large class of algebraic curves, such
divisors can be found in refs. \cite{acsi}. Analogously, the metric \ref
{nondegthree} becomes on a general algebraic curve: 
\begin{equation}
\widetilde{g}_{z\overline{z}}dzd\overline{z}=\frac{e^{\left| F_y \right|^2}}{%
e^{\left|F_y\right|^2}-1}\left[ 1+z\overline{z}\right] ^\beta dzd\overline{z}
\end{equation}
for suitable values of $\beta$. The results of Section 3 allows us to write
explicitly the Lagrangians of many field theories on algebraic curves. For
instance, let us write the action for the scalar fields $\varphi(z,\bar
z;y^{\tilde l}(z),\overline{y^{\tilde l}(z)}$ with mass $\mu$: 
\begin{equation}
S=\sum_{\tilde l=0}^{n-1} \int_{{\bf CP}_1}d^2z\left[ \frac 12\left(
d_z\varphi ^{(\widetilde{l})}d_{\overline{z}}\varphi ^{(\widetilde{l})}+\mu
^2g_{z\overline{z}}^{(\widetilde{l})}(\varphi ^{(\widetilde{l})})^2 +\lambda
_1R_{z\overline{z}}^{(\widetilde{l})}(\varphi ^{(\widetilde{l})})^2\right)
+\lambda _2R_{z\overline{z}}^{(\widetilde{l})}\varphi ^{(\widetilde{l}%
)}\right]  \label{esse}
\end{equation}
Here $g_{z\overline{z}}$ is a general metric on $\Sigma $ with Ricci tensor $%
R_{z\overline{z}}$, $\tilde l=0,\cdots ,n-1$ is the branch index and $%
\lambda _1,\lambda _2$ represent real parameters. Explicit examples of
metrics and curvatures have been given above and in Section 3. Let us notice
that the integration in the right hand side of eq. \ref{esse} is over {\bf CP%
}$_1$ after applying the Poincar\'e--Lelong equation but the integrand is
multivalued. Moreover, $d_z$ and $d_{\overline{z}}$ are total derivatives
with respect to the variables $z$ and $\overline{z}$. Total derivatives are
used to remember that the fields $\varphi $ depend on $z,\overline{z}$ also
through the functions $y(z),\overline{y(z)}$. Deriving the action \ref{esse}
with respect to the field $\varphi $ in a given branch $l$, we find the
equation of motion of the scalar fields:

\begin{equation}
-d_zd_{\overline{z}}\varphi ^{(l)}+\left( \mu ^2g_{z\overline{z}%
}^{(l)}+\lambda _1R_{z\overline{z}}^{(l)}\right) \varphi ^{(l)}+\lambda
_2R_{z\overline{z}}^{(l)}=0  \label{sfeq}
\end{equation}

where 
\begin{equation}
d_zd_{\overline{z}}\varphi =\partial _z\partial _{\overline{z}}\varphi
+\partial_z\partial_{\bar y}\varphi\frac{d\bar y}{d\bar z}+
\partial_y\partial_{\bar z}\varphi\frac{dy}{dz} +\partial _y\partial _{%
\overline{y}}\varphi \left| \frac{dy}{dz}\right| ^2  \label{totder}
\end{equation}

Locally and far from the branch points, it is possible to solve \ref{sfeq}
with the standard methods of the theory of partial differential equations on
the complex plane. In the case of the $Z_n$ symmetric curves we can even
choose singlevalued metrics and curvatures, so that the coefficients
appearing in the differential equation \ref{sfeq} are singlevalued.
Nevertheless, any local solution derived in this way is in general
multivalued and needs to be analytically continued in order to extend it
over the whole algebraic curve. Despite of the difficulties that may arise
in the analytic continuation, the possibility of transforming differential
equations on a Riemann surface in differential equations on the sphere is
remarkable. Moreover, numerical calculations are allowed due the
explicitness which is intrinsic in the representation of Riemann surfaces in
terms $n-$sheeted coverings of the complex sphere.

\section{ACKNOWLEDGEMENTS}

The author would like to thank J. Sobczyk for participating in the
preliminary stages of this work and for many helpful discussions.
This work has been in part supported by the European Community,
TMR grant ERB4001GT951315.

\vfill\eject

\begin{center}
FIGURE CAPTIONS
\end{center}

\vspace{1cm}

\begin{description}
\item[\tt 1)]  A possible set of branch cuts on the complex sphere for the $%
Z_n$ symmetric algebry curves. The cuts appear symmetrically on the sheets
composing the curve.\\

\item[\tt 2)]  A possible covering of the curve $\Sigma$ in two sets $%
\Sigma_N$ and $\Sigma_N$. Only the part of the contour $\gamma$ which lies
on the $i-$th sheet is showed.\\

\item[\tt 3)]  An alternative form of the two sets $\Sigma_N$ and $\Sigma_S$%
. $\Sigma_S$ is disconnected in $n$ pieces lying on the different sheets. In
the figure only the piece belonging to the $i-$th sheet has been given.
\end{description}

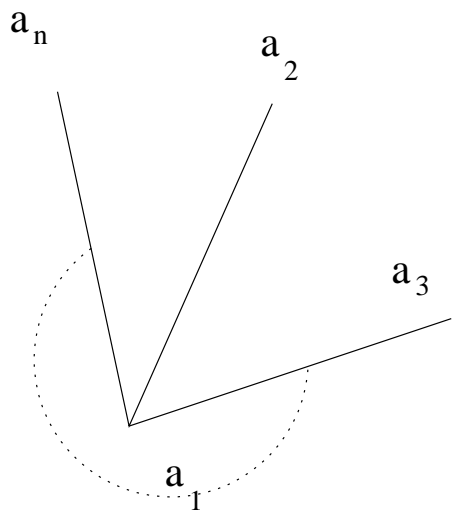
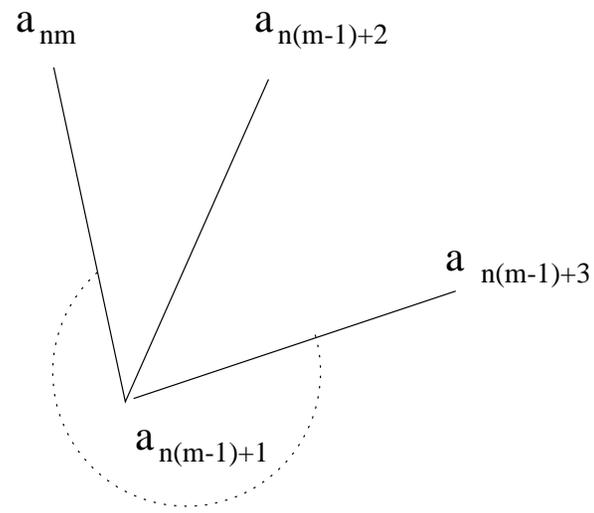
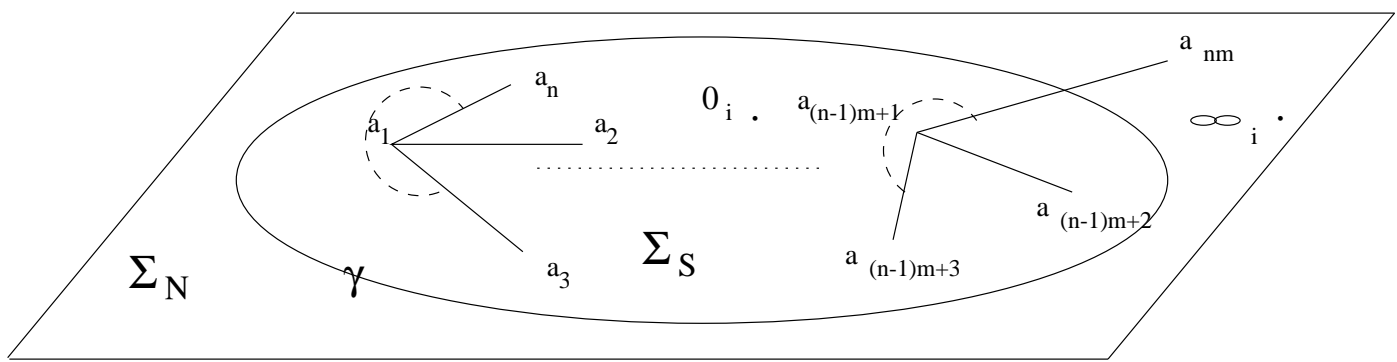
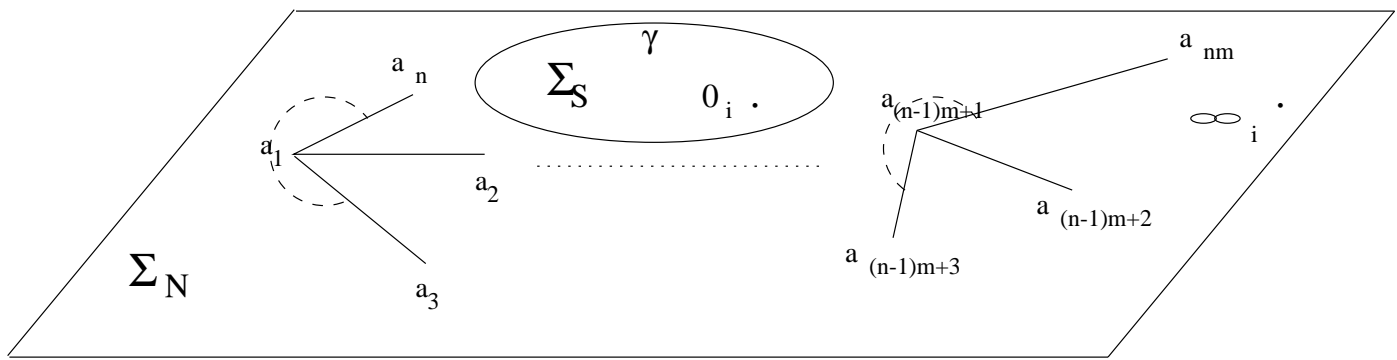
\end{document}